\documentclass[a4paper,11pt]{article}
\usepackage{pos}
\usepackage{subcaption}

\title{Light to Heavy, Brief to Eternal: \\ An Axion for Every Occasion (in the Early Universe)}
\ShortTitle{An Axion for Every Occasion (in the Early Universe)}

\author*[a,b]{Francesco D'Eramo}

\affiliation[a]{Dipartimento di Fisica e Astronomia,  Università degli Studi di Padova,\\ Via Marzolo 8, 35131 Padova, Italy}

\affiliation[b]{Istituto Nazionale di Fisica Nucleare (INFN), Sezione di Padova,\\ Via Marzolo 8, 35131 Padova, Italy}

\emailAdd{francesco.deramo@pd.infn.it}

\abstract{The early universe grants access to energy scales far beyond those achievable in terrestrial experiments and allows unstable Standard Model particles to play an active dynamical role. In this contribution, we focus on recent studies aimed at quantifying the potential of the early universe to probe the properties and interactions of axions. The discussion is organized around four classes of axion scenarios, ordered from long to short lifetimes: (i) stable or long-lived axions contributing to dark radiation; (ii) stable or long-lived axions produced out-of-equilibrium and constituting dark matter; (iii) metastable axions whose decays inject energy into the primordial plasma and leave observable signatures in the global 21 cm signal; and (iv) very short-lived axions that act only as portals to additional degrees of freedom. Together, these scenarios highlight the interplay between axion phenomenology and early universe cosmology and demonstrate the potential of cosmological data to probe axions over a broad range of masses and lifetimes.}

\FullConference{Proceedings of the Corfu Summer Institute 2025 "School and Workshops on Elementary Particle Physics and Gravity" (CORFU2025)\\
27 April - 28 September, 2025\\
Corfu, Greece\\}

\tableofcontents

\begin{document}
\maketitle

\section{Introduction}
\label{sec:intro}

Axions are hypothetical degrees of freedom arising in extensions of the Standard Model (SM) and are strongly motivated by top-down considerations. Foremost among these is the Peccei-Quinn (PQ) mechanism~\cite{Peccei:1977hh,Peccei:1977ur}, which provides an elegant solution to the strong CP problem and predicts an axion as its only low energy remnant~\cite{Wilczek:1977pj,Weinberg:1977ma}. Such particles possess a shift symmetry inherited from the underlying ultraviolet theory and are therefore naturally light. Their mass can be understood as arising from a soft breaking of this symmetry at an appropriate energy scale. Throughout this work, the axion mass $m_a$ is treated as a free parameter, corresponding to such a soft breaking.

Moreover, the Nambu-Goldstone nature of axions implies that their interactions are generically suppressed by the breaking scale of the global symmetry responsible for their existence. As a result, axions are not only naturally light but also naturally weakly coupled. In particular, axions couple to fermions through shift-symmetric derivative interactions, while their couplings to gauge bosons arise from anomalous terms~\cite{Georgi:1986df}. To illustrate the general framework, axion interactions can be written schematically as
\begin{equation}
\mathcal{L}^{\rm int}_a =
\frac{a}{8\pi f_a} \sum_V C_V V^{\mu\nu}\widetilde{V}_{\mu\nu}
+ \frac{\partial_\mu a}{f_a} \sum_\psi C_\psi \bar{\psi}\gamma^\mu\gamma_5\psi \, .
\label{eq:Laint}
\end{equation}
Here we include both anomalous couplings to gauge bosons $V$ and derivative couplings to fermions $\psi$, retaining only axion interactions up to dimension five operators suppressed by the axion decay constant $f_a$. Once axions couple to visible-sector fields through the interactions in Eq.~\eqref{eq:Laint}, their presence has unavoidable consequences for the evolution of the universe. These consequences can be quantified in terms of the effective couplings appearing in the Lagrangian, $C_X / f_a$ with $X = V, \psi$. Doing so relies on a rather mild and reasonable assumption, which we discuss next.

Within the standard cosmological model, extrapolating the expansion history backward in time using known physical laws yields a coherent and robust picture. At sufficiently early times, the energy density of the universe was dominated by a relativistic plasma of SM particles in thermal equilibrium. This framework is supported directly by observations of the Cosmic Microwave Background (CMB), originating at the surface of last scattering when the universe was approximately $380{,}000$ years old, and indirectly by the remarkable agreement between the predicted and observed light-element abundances from Big Bang Nucleosynthesis (BBN). The earliest process probed by BBN is neutron decoupling, which occurs when the universe is about one second old, at temperatures of order MeV, well below the masses of most SM particles.

In this contribution, we study the cosmological implications of an axion coupled to SM fields under the assumption that the radiation-dominated phase of the universe extends at least up to temperatures of order the TeV scale. Equivalently, we assume that at some stage during cosmic expansion all SM degrees of freedom were relativistic and in thermal equilibrium at temperatures above the weak scale. Although this description must ultimately break down, for instance during reheating after inflation, we assume that radiation domination persists at least up to the TeV scale. This assumption is reasonable, as it corresponds to extrapolating the BBN snapshot by only a few orders of magnitude in the plasma temperature.

Within this framework, cosmological history can be viewed as a timeline that serves as a reference for discussing the cosmological role of axions. As we will show through explicit examples, axions are generically produced via thermal scatterings and/or decays of SM particles. Their production is therefore not optional but rather \textit{unavoidable}, and must be consistently accounted for and quantified. This notion of \textit{inevitability} has already proven to be a powerful tool for deriving cosmological constraints on bosonic particles in a wide range of settings~\cite{Langhoff:2022bij,DEramo:2024lsk}.

Axions produced in the early universe do not persist indefinitely. Although they are naturally light and weakly coupled, no symmetry protects their stability, and they are therefore generically unstable. This remains true even when the SM particles to which axions couple through the interactions in Eq.~\eqref{eq:Laint} are heavier than the axion itself. Beyond tree level, and at sufficiently high order in perturbation theory, a decay channel into two photons is always present. As a consequence, for any finite coupling to the visible-sector fields, axions ultimately decay back into SM particles.

The timing of these decays plays a crucial role in determining their cosmological impact. Depending on when they occur, the associated energy injection can influence different stages of cosmic evolution. Axion phenomenology therefore depends sensitively on both the axion mass and lifetime, ranging from effectively stable relics to rapidly decaying states. As we will see, in several cases axions leave observable imprints in the early universe, making cosmology a powerful probe of axion parameter space that is often inaccessible to terrestrial experiments.

These two basic features, thermal production in the early universe and eventual decay at late times, define the natural beginning and end points of the axion cosmological history and provide the organizing principle of this manuscript. In Sec.~\ref{sec:origends}, we show how this perspective motivates a classification of the axion parameter space into three regions associated with distinct cosmological imprints. We then present explicit realizations of these scenarios, discussing dark radiation in Sec.~\ref{sec:DR}, dark matter in Sec.~\ref{sec:DM}, energy injection affecting the global 21 cm signal in Sec.~\ref{sec:21cm}, and a novel realization of the axion portal in Sec.~\ref{sec:portal}. Finally, we summarize our findings in Sec.~\ref{sec:conclusions}.

\section{Unavoidable Origins, Inevitable Ends}
\label{sec:origends}

In this section, we define the general setup more concretely by focusing on a particularly instructive example, namely the case in which the axion couples predominantly to photons. For any finite axion mass, the decay into two photons, $a \rightarrow \gamma\gamma$, is kinematically allowed and provides a clean benchmark for understanding the role of lifetimes in axion phenomenology. Even if such a coupling is absent at tree level, it is typically generated radiatively and can therefore dominate the axion decay channels. For these reasons, the scenario in which the axion decays predominantly into photons represents a robust and well motivated case to study.

Up to numerical factors, the corresponding lifetime scales as $\tau_{a \rightarrow \gamma\gamma} = \Gamma_{a \rightarrow \gamma\gamma}^{-1} \propto ( f_a / C_\gamma )^2 m_a^{-3}$. This simple scaling illustrates that the axion lifetime can span many orders of magnitude, ranging from extremely rapid decays occurring well before the first second after the Big Bang to cosmologically long lifetimes far exceeding the current age of the universe. The lifetime is controlled by two parameters, the axion mass and its coupling strength. Heavier axions and larger couplings lead to shorter lifetimes, while lighter and more weakly-coupled axions can remain effectively stable on cosmological timescales.

The example discussed above can be readily generalized to axion decays into generic SM final states, $a \rightarrow XX$. Here, $X$ may represent either a gauge boson or a fermion, and the interaction governing the dominant decay channel can likewise arise radiatively. The dependence of the axion lifetime on the decay constant and the mass varies according to whether the decay occurs at tree level and on the nature of the final state. Nevertheless, the qualitative conclusion remains unchanged: heavier and more strongly coupled axions decay more rapidly.

\begin{figure}
    \centering
    \includegraphics[width=\textwidth]{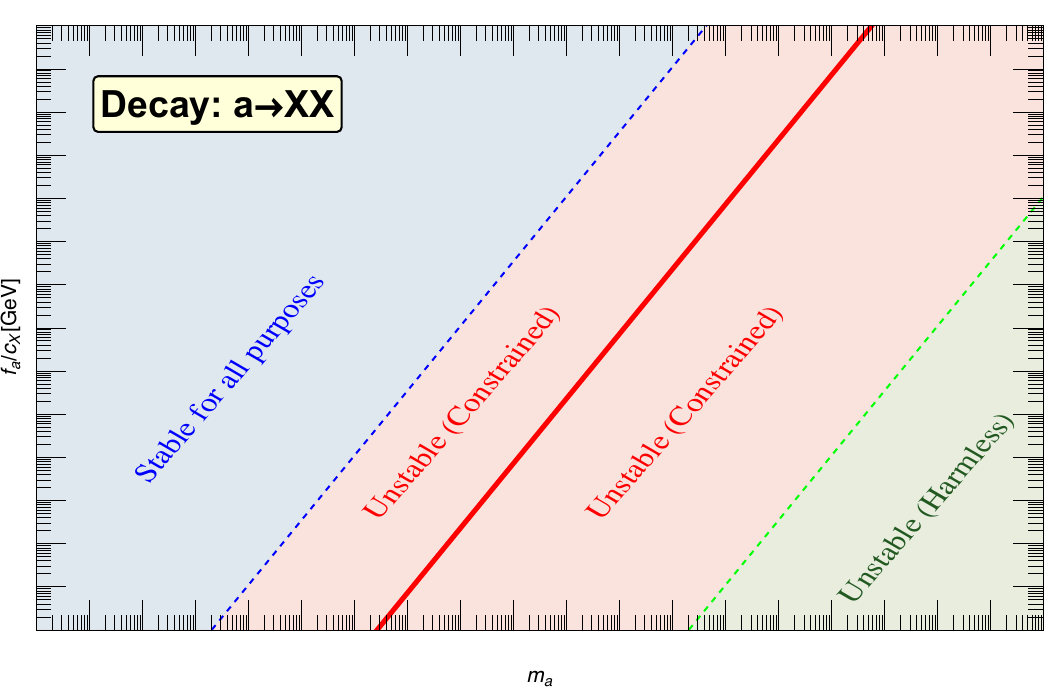}
    \caption{Schematic illustration of the axion parameter space in the plane spanned by the axion mass $m_a$ and the inverse coupling $f_a / C_X$ to a generic visible-sector particle $X$. Lines of constant axion lifetime for the decay $a \rightarrow XX$ partition the parameter space into three qualitatively distinct regions. From right to left, these correspond to short-lived axions that decay before BBN, shown in green; axions with intermediate lifetimes whose decays inject energy and can affect cosmological observables, shown in red, with the solid line indicating the current age of the universe; and effectively stable axions that survive to late times and may contribute to dark matter or dark radiation, shown in blue.}
    \label{fig:axion_lifetime_plane}
\end{figure}

A convenient way to visualize the relevant parameter space is the plane spanned by the axion mass and the inverse coupling $f_a / C_X$. This parameter space is shown in Fig.~\ref{fig:axion_lifetime_plane}, where lines of constant lifetime partition the plane into three qualitatively distinct regions. For sufficiently large couplings and/or large masses, shown in green, axions are very short lived and decay prior to BBN, leaving little impact on late time cosmological observables. For intermediate lifetimes, shown in red and corresponding to decays occurring after BBN but before approximately $10^{27}$ seconds, the associated energy injection can modify the cosmological history and is therefore subject to stringent constraints. Finally, for sufficiently small couplings and/or small masses, shown in blue, the axion lifetime becomes extremely long, rendering the axion effectively stable on cosmological timescales and allowing it to contribute as dark matter or dark radiation.

This simple example illustrates a general principle: the axion lifetime, determined by its mass and coupling, largely dictates its cosmological phenomenology. The parameter space shown in Fig.~\ref{fig:axion_lifetime_plane}, organized in terms of mass, coupling, and lifetime, provides a useful map for classifying cosmological scenarios. Short-lived axions decay before BBN, axions with intermediate lifetimes decay at cosmologically sensitive epochs, and effectively stable axions persist to the present day.

Adopting this perspective, the remainder of the discussion can be naturally organized according to these regions. In the weakly-coupled and effectively stable regime lie scenarios associated with dark radiation and dark matter. In the intermediate lifetime region, axion decays can leave observable signatures in the 21 cm signal. Finally, in the strongly-coupled and short-lived regime, axions act as portals to dark sectors. Taken together, these examples demonstrate how the mass-coupling plane, structured by lifetime, captures the essential phenomenology across a broad range of axion models.

\section{Stable Axions I: Dark Radiation}
\label{sec:DR}

We begin by considering the region on the left side of the parameter space shown in Fig.~\ref{fig:axion_lifetime_plane}, focusing on relatively light axion masses. In particular, we concentrate on thermally produced axions that contribute as an additional relativistic component of the cosmic energy budget, commonly referred to as dark radiation. The presence of dark radiation modifies the expansion history of the universe and leaves characteristic imprints on cosmological observables, most notably on the inferred radiation density and the corresponding evolution of the Hubble rate~\cite{Brust:2013ova}. These signatures provide a powerful window into otherwise inaccessible regions of axion parameter space~\cite{Baumann:2016wac,Ferreira:2018vjj,DEramo:2018vss,Arias-Aragon:2020qtn,Arias-Aragon:2020shv,Green:2021hjh}. As discussed below, this requirement restricts our attention to the region satisfying $m_a \lesssim 0.1\,\mathrm{eV}$.

Assuming an inflationary reheating scenario in which axions are not produced directly from inflaton decays, scatterings and/or decays of particles in the thermal bath can nevertheless lead to final states containing axions. These processes arise as a natural consequence of the interactions in Eq.~\eqref{eq:Laint} and of the presence of a thermal bath populated by SM particles, which are abundant since they are relativistic and therefore not Maxwell--Boltzmann suppressed. The relevant question is not whether axion production occurs, but rather how efficiently it proceeds and what final abundance is generated. In particular, thermally produced axions may reach thermal equilibrium and subsequently decouple from the plasma through a \textit{freeze-out} process. Alternatively, if the interactions are sufficiently weak, axions may never thermalize, and their population is instead generated via the \textit{freeze-in} mechanism.

At some point in the expansion history, regardless of whether they ever reached thermalization at earlier times, the thermally produced axions propagate along geodesics of the expanding universe. In this regime, their energy density $\rho_a$ redshifts with the scale factor $a$ as $\rho_a \propto a^{-4}$ because they are relativistic. They will not be relativistic forever since their energy and momentum are subject to the cosmological redshift, and eventually their kinetic energy will be of the order of the mass energy. Therefore, the transition between the relativistic and the non-relativistic regime happens at a moment that depends on the axion mass $m_a$. 

To estimate when this transition occurs, it is crucial to keep in mind the thermal origin of axions. Although axions are not produced monochromatically but instead with a momentum distribution resulting from the thermal ensemble of incoming states, they originate from microscopic processes involving bath particles with typical energies of order the plasma temperature $T$. Once axions decouple from the thermal bath, their energy redshifts as $E_a \propto a^{-1}$, while the temperature of the plasma evolves according to entropy conservation, $g_{\star s}(T)^{1/3} T a = \mathrm{const}$. Up to corrections associated with the variation of the effective number of entropic degrees of freedom $g_{\star s}(T)$, the axion energy and the plasma temperature therefore decrease in an almost identical manner. As a consequence, the transition between the relativistic and non-relativistic regimes for axions occurs when the plasma temperature drops to $T_{\rm NR} \simeq m_a$. In other words, as long as the bath temperature remains above $m_a$, thermally produced axions contribute to the radiation energy density.

Historically, the presence of dark radiation has been parameterized in terms of an effective number of additional neutrino species, $\Delta N_{\rm eff}$. This quantity is probed at two crucial epochs in the expansion history of the universe: BBN~\cite{Yeh:2022heq,Schoneberg:2024ifp} and the CMB formation~\cite{Planck:2018vyg,ACT:2020gnv}. In light of the significant improvements expected in future CMB determinations of $\Delta N_{\rm eff}$~\cite{SimonsObservatory:2018koc,Abazajian:2019eic,CMB-S4:2022ght}, we focus on regions of parameter space in which thermally produced axions remain relativistic also at the time of CMB formation. This requirement translates into the condition $m_a \lesssim 0.1\,\mathrm{eV}$.

Evaluating $\Delta N_{\rm eff}$ requires determining the axion energy density. Below the electron positron annihilation threshold at $511\,\mathrm{keV}$, there are no further changes in the effective number of entropic degrees of freedom $g_{\star s}(T)$. From this point onward, the ratio between the axion and photon energy densities, $\rho_a / \rho_\gamma$, therefore remains constant. This ratio can be directly related to the effective number of additional neutrino species as
\begin{equation}
\Delta N_{\rm eff} = \frac{8}{7} \left( \frac{11}{4} \right)^{4/3} \frac{\rho_a}{\rho_\gamma} \ .
\end{equation}
The task at hand is therefore to evaluate the asymptotic ratio $\rho_a / \rho_\gamma$ as a function of the Lagrangian parameters appearing in Eq.~\eqref{eq:Laint}.

The work presented in Ref.~\cite{DEramo:2023nzt} developed a rigorous framework for computing dark radiation production in the early universe by tracking the full phase space distribution of a generic light particle produced through scatterings and/or decays of degrees of freedom in the primordial thermal bath. Solving the Boltzmann equation in momentum space is the appropriate approach for accurately following the evolution of the axion population, and goes beyond commonly employed simplified treatments that assume instantaneous decoupling or track only the number density evolution under the assumption of a thermal momentum distribution. This framework further allows for the consistent inclusion of quantum statistical effects as well as energy exchange between the dark sector and the SM. Moreover, it enables a reliable determination of the dark radiation contribution even in scenarios where the light particle never reaches full thermal equilibrium or decouples gradually from the thermal bath. A comparison between the full momentum space evaluation of $\Delta N_{\rm eff}$ and commonly used approximate methods, performed in a largely model-independent manner, shows that analyses based solely on the evolution of the number density are sufficient to confront current observational bounds. However, such approximations are generally inadequate to fully exploit the sensitivity of future cosmological measurements~\cite{DEramo:2023nzt} .

\begin{figure}[t]
    \centering
    \begin{subfigure}[b]{0.45\textwidth}
        \centering
        \includegraphics[width=\textwidth]{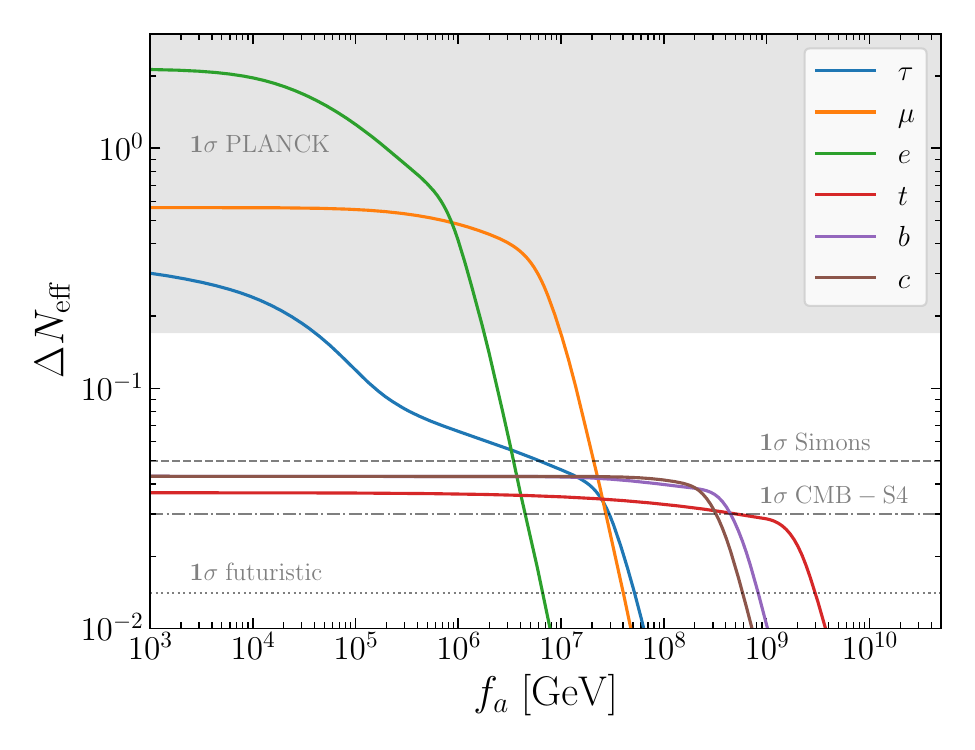}
        \caption{Predicted $\Delta N_{\rm eff}$ from a full phase space analysis as a function of the axion decay constant $f_a$, for axion couplings to charged leptons and heavy quarks, with a single Wilson coefficient set to $c_\psi = 1$ in each case.}
        \label{fig:axionDRa}
    \end{subfigure}
    \hfill
    \begin{subfigure}[b]{0.45\textwidth}
        \centering
        \includegraphics[width=\textwidth]{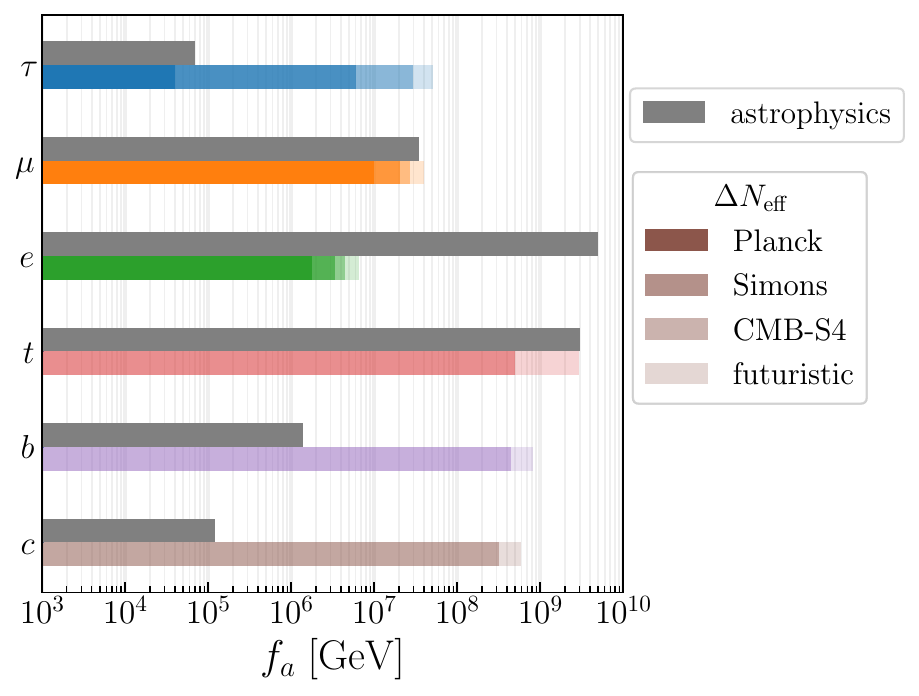}
        \caption{Current and projected cosmological constraints on the axion decay constant obtained from a full phase space analysis. Existing astrophysical bounds are shown as gray bands for comparison.}
        \label{fig:axionDRb}
    \end{subfigure}
    \caption{Axion dark radiation arising from couplings to SM fermions. The predicted contributions to $\Delta N_{\rm eff}$ are obtained from a full phase space analysis that consistently tracks the momentum distribution of thermally produced axions. The results shown here are taken from Ref.~\cite{DEramo:2024jhn}.}
    \label{fig:axionDR}
\end{figure}

In Fig.~\ref{fig:axionDR} we present results from the recent analysis of Ref.~\cite{DEramo:2024jhn}. The focus is on flavor-diagonal axion couplings%
\footnote{Flavor-violating axion couplings can also give rise to cosmological signatures that lead to constraints complementary to those obtained from astrophysics and terrestrial experiments~\cite{Baumann:2016wac,Green:2021hjh,DEramo:2021usm}. A phase space analysis for flavor-violating axion couplings to leptons can be found in Ref.~\cite{Badziak:2024qjg}.}
to charged leptons and heavy quarks. The corresponding dark radiation predictions are obtained using the momentum space framework developed in Ref.~\cite{DEramo:2023nzt}.

The left panel in Fig.~\ref{fig:axionDRa} shows the dependence of $\Delta N_{\rm eff}$ on the axion decay constant $f_a$, with the dimensionless Wilson coefficient of the relevant operator set to unity. The predictions for charged leptons are theoretically very robust, as leptons do not carry color and are therefore unaffected by uncertainties associated with QCD confinement. The top quark is sufficiently heavy compared to the QCD scale that the corresponding results are also very solid. By contrast, the bottom quark mass lies close to the confinement scale. As shown in Ref.~\cite{DEramo:2024jhn}, the predictions in this case remain reliable, while the curve associated with the charm quark is significantly more sensitive to the choice of the temperature at which the Boltzmann evolution is terminated.

The right panel in Fig.~\ref{fig:axionDRb} shows current and projected cosmological bounds on the axion decay constant $f_a$ for the same physical setup. For comparison, existing astrophysical bounds are also displayed in the same figure. Remarkably, observations of the CMB already provide, and are expected to provide even more stringent constraints in the future, the strongest limits on axion couplings to some SM fermions.

To summarize, the unavoidable production of axions in the early universe provides an important avenue for probing axion couplings to visible-sector fields. The results presented in Fig.~\ref{fig:axionDR} represent state-of-the-art predictions for axion interactions with SM fermions, based on a full momentum-space treatment without additional simplifying assumptions. Axion interactions with gauge bosons are also of significant interest, and couplings to gluons in particular are strongly motivated by their connection to the strong CP problem. In this case, however, additional theoretical challenges arise due to the infrared behavior of diagrams involving massless gauge bosons exchanged in the $t$-channel, which require a proper treatment within thermal field theory~\cite{Graf:2010tv,Salvio:2013iaa,DEramo:2021psx,DEramo:2021lgb,Bouzoud:2024bom}. Overcoming these limitations is therefore a key objective. In particular, extending the phase-space analysis to axion couplings to gauge bosons, and to gluons in particular, will be essential to fully exploit the upcoming wealth of cosmological data.

\section{Stable Axions II: Dark Matter}
\label{sec:DM}

The second case we consider also lies in the blue region of Fig.~\ref{fig:axion_lifetime_plane}, but corresponds to larger values of the axion mass. In particular, we are interested in determining when thermally produced axions can contribute to the observed dark matter abundance. For the reasons discussed in the previous paragraph, this requires axion masses above the sub-eV scale. Once this threshold is exceeded, axions no longer contribute to the radiation energy density at the time of CMB formation, as the cosmological redshift renders them non-relativistic.

Let's assume for a moment that the coupling strength is large enough to bring axions in thermal equilibrium. The story in this case is very similar to the one of massive SM neutrinos. They are present with the same relativistic abundance, up to order one factors, and therefore they induce a free-streaming that is constrained by large scale structure. Bounds have been put in this context on the axion coupling to pions~\cite{Ferreira:2020bpb,Notari:2022ffe,Bianchini:2023ubu}, photons~\cite{Caloni:2022uya}, leptons~\cite{Badziak:2025mkt}, and UV complete frameworks for the QCD axion such as the KSVZ and the DFSZ realizations~\cite{DEramo:2022nvb}. They all constraint the ballpark mass range $m_a \lesssim 0.2 \, {\rm eV}$. Even if we saturate this inequality, the analogy with SM neutrinos would extend: thermally produced massive axions would constitute a sub-dominant hot dark matter component. 

We are interested here in scenarios in which axions produced from the thermal bath account for the entire dark matter abundance. This requires sufficiently large axion masses and, once the Cowsik--McClelland bound is taken into account~\cite{Cowsik:1972gh}, excludes production via thermal freeze-out in the early universe. We therefore focus on axion couplings that are so feeble that thermal equilibrium is never attained, yet still large enough to yield a non-negligible axion abundance. In this regime, axions are produced through rare scatterings and/or decays of particles in the thermal bath and are injected into the early universe. The inverse processes are always negligible due to the small axion abundance, and the produced axions subsequently propagate along geodesics and persist until the present day. In other words, we focus on the freeze-in production of axion dark matter.

The freeze-in scenario~\cite{Hall:2009bx} is often regarded as a nightmare for experimental searches, due to the extremely small couplings involved. Nevertheless, not all freeze-in scenarios are entirely hopeless, and some can still lead to observable signatures. One of the earliest proposals in this direction was the possibility of detecting displaced events at particle colliders~\cite{Co:2015pka,Evans:2016zau,DEramo:2017ecx,Calibbi:2018fqf,Belanger:2018sti,Bae:2020dwf,Calibbi:2021fld}. 

Here, we focus on a different and characteristic imprint of freeze-in production, namely the erasure of cosmological perturbations at small scales due to dark matter free-streaming~\cite{McDonald:2015ljz,Roland:2016gli,Heeck:2017xbu,Bae:2017dpt,Kamada:2019kpe,DEramo:2020gpr}. The focus in this contribution is on the results of Ref.~\cite{DEramo:2025jsb} which studied axion freeze-in production via the derivative couplings to fermions\footnote{The impact on structure formation of axion freeze-in production via coupling to photons can be found in Refs.~\cite{Baumholzer:2020hvx,Becker:2025yvb}.} appearing in Eq.~\eqref{eq:Laint}. As discussed for the dark radiation case in the previous section, only axion couplings to charged leptons and heavy quarks were considered.

Before turning to the specific case of axions, it is useful to place the discussion in a broader context that applies to a wide class of dark matter candidates. Indeed, the analysis can be carried out in a largely model-independent manner. This approach was adopted in Ref.~\cite{DEramo:2025jsb}, building on the results of Ref.~\cite{DEramo:2020gpr}, which established a correspondence between freeze-in dark matter and warm dark matter (WDM). In particular, it was shown that the matter power spectrum associated with freeze-in dark matter features a sharp suppression of small-scale structure, closely resembling that of WDM. As a result, a given freeze-in scenario can be mapped onto an equivalent WDM model by selecting an appropriate value of the WDM mass. 

The key quantity that enables this mapping is the dark matter root mean square velocity~\cite{Kamada:2019kpe}. The procedure for deriving a mass bound on freeze-in dark matter proceeds as follows. First, the full phase space distribution of dark matter is obtained by solving the Boltzmann equation in momentum space, including all relevant collision terms. Once the distribution is known, the root mean square velocity of dark matter is computed and compared to the maximum allowed warmness inferred for warm dark matter. This comparison yields a lower bound on the mass of freeze-in dark matter~\cite{DEramo:2025jsb}
\begin{equation}
\label{eq:warmness_map}
m_\chi^{\rm min} = 22\ {\rm keV}
\left( \frac{m_{\rm WDM}^{\rm min}}{6\ {\rm keV}} \right)^{4/3}
\left( \frac{\sigma_q}{3.6} \right)
\left( \frac{106.75}{g_{\star s}(T_P)} \right)^{1/3} \, ,
\end{equation}
where $\sigma_q$ denotes the second moment of the phase space distribution expressed in comoving variables, and $T_P$ is the characteristic production temperature.\footnote{It is worth emphasizing that the mass bound in Eq.~\eqref{eq:warmness_map} is independent of the arbitrary choice of the production temperature $T_P$. While the second moment $\sigma_q$ does depend on $T_P$, this is exactly compensated by the factor $g_{\star s}(T_P)$.}

\begin{figure}
    \centering
    \includegraphics[width=\textwidth]{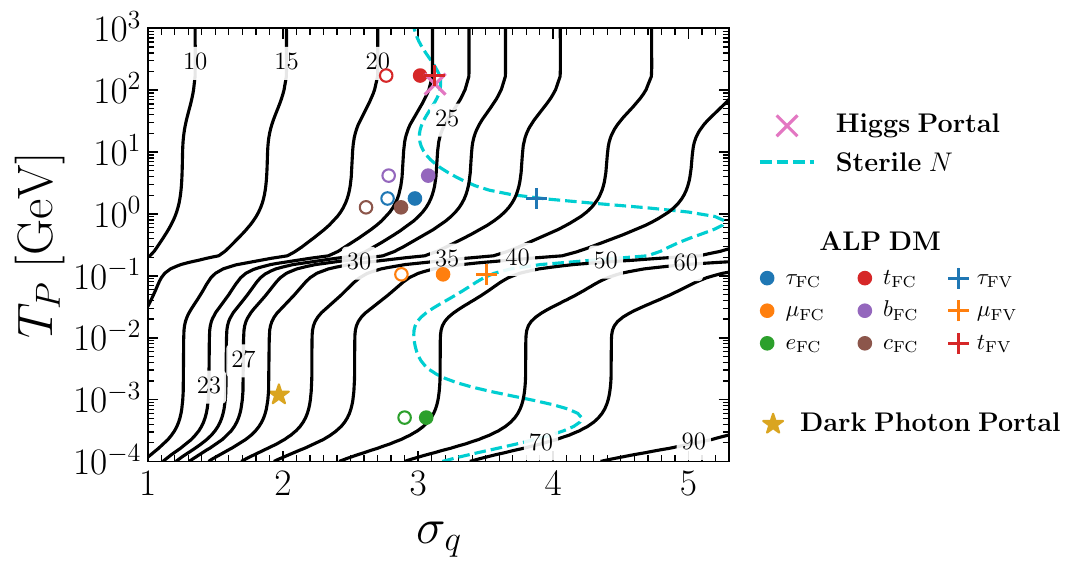}
    \caption{Compact and model-independent visualization of mass bounds on freeze-in dark matter. The horizontal and vertical axes correspond to the second moment of the momentum distribution, $\sigma_q$, and the characteristic production temperature $T_P$. Black isocontours indicate the minimum allowed dark matter mass from small-scale structure constraints, obtained by rescaling to a reference warm dark matter mass of $6.8\,\mathrm{keV}$. Superimposed on these model-independent contours are benchmark realizations, with $T_P$ chosen to reflect the characteristic mass or energy scale of the dominant production process. Figure from Ref.~\cite{DEramo:2025jsb}.}
    \label{fig:DM}
\end{figure}

The plot shown in Fig.~\ref{fig:DM} provides a compact and model-independent summary of the mass bounds that can be obtained using this procedure. In the $(\sigma_q, T_P)$ plane, the black isocontours identify the minimum allowed dark matter mass once the WDM bound is fixed to $6.8\,\mathrm{keV}$. In the same plane, several benchmark models are indicated, including axions produced through interactions with SM fermions. As can be seen from these results, the resulting mass bounds lie in the approximate range between $20$ and $50\,\mathrm{keV}$. The most stringent bounds arise from couplings to lighter fermions. This behavior can be understood by noting that axion production in these cases is infrared dominated, so that production occurs at later times, when the effective number of entropic degrees of freedom $g_{\star s}(T_P)$ is smaller. We also note that the case of axions coupled to electrons is not viable, since radiative corrections induce an axion photon coupling that leads to the decay $a \rightarrow \gamma\gamma$ with a lifetime excluded by X-ray line searches.

\section{Unstable Axions: 21 cm Imprints}
\label{sec:21cm}

We now turn to the red region of the parameter space shown in Fig.~\ref{fig:axion_lifetime_plane}, where the axion lifetime is too short for the particle to be considered effectively stable, yet remains longer than one second. In this regime, axion decays can significantly affect cosmological evolution and potentially disrupt the successful predictions of the standard cosmological model. From a more optimistic standpoint, however, the same decay products can give rise to observable signatures absent from the standard cosmological framework, thereby providing a window onto new physics beyond the SM.

This contribution examines axions produced in the early universe that inject energy into the primordial plasma at later times through their decays. We focus on the resulting imprints on the 21\,cm line of neutral hydrogen. The 21\,cm observable probes the thermal and ionization history of the cosmic dark ages, well before the formation of the first luminous sources. Measurements of its absorption or emission relative to the CMB are therefore sensitive to subtle deviations from standard cosmological evolution induced by exotic energy injection. Although a definitive detection of the global 21\,cm signal has not yet been achieved, a broad experimental program is currently underway worldwide, encompassing both global-signal experiments and interferometric arrays.

In this context, Ref.~\cite{Cima:2025zmc} has recently investigated the implications of non-minimal dark sectors within a model-independent framework. Two representative scenarios were considered. In the first, the dark sector contains a cold dark matter component that reproduces the observed relic abundance, together with an additional metastable state. In the second, the dark sector consists of two nearly degenerate particle species, with the heavier one being metastable. In both scenarios, decays of dark sector states inject energy into the primordial plasma and can significantly modify the evolution of the 21\,cm signal. Overall, this approach highlights the remarkable sensitivity of 21\,cm cosmology to new metastable particles that have otherwise disappeared from the present day universe.

Our focus here is on the global 21\,cm signal associated with the onset of star formation, which occurs at typical redshift values $z_t \simeq 15$. Within the $\Lambda$CDM model, this epoch is expected to produce an absorption feature in the brightness temperature whose amplitude depends sensitively on astrophysical assumptions. This dependence reflects the large uncertainties associated with the details of early star formation. In the extreme limit where astrophysical heating is completely neglected, the brightness temperature reaches a minimum value\footnote{Here, the minimum value corresponds to the largest signal in absolute magnitude, since the brightness temperature is expected to be negative at these redshifts.} that defines the maximal amplitude of the signal, $\mathcal{A}_{\Lambda{\rm CDM}}^{\rm max}(z)$, as a function of redshift. Any additional exotic energy injection, such as that arising from the decays of metastable particles, increases the gas temperature and consequently makes the signal less negative. In the presence of such effects, the maximal amplitude is therefore shifted to a larger value, denoted $\mathcal{A}^{\rm max}_\chi(z)$. If a signal is detected, we denote the corresponding measured amplitude by $\mathcal{T}(z)$, leading to the chain of inequalities $\mathcal{A}_{\Lambda{\rm CDM}}^{\rm max}(z) < \mathcal{A}^{\rm max}_\chi(z) < \mathcal{T}(z)$. The first two quantities can be computed theoretically, while the third awaits experimental determination and can presently be replaced by a reference threshold value. This strategy is deliberately conservative, as it attributes all heating of the intergalactic medium to exotic energy injection while completely neglecting the astrophysical heating expected within the $\Lambda$CDM model.

\begin{figure}[t]
    \centering
    \begin{subfigure}[b]{0.45\textwidth}
        \centering
        \includegraphics[width=\textwidth]{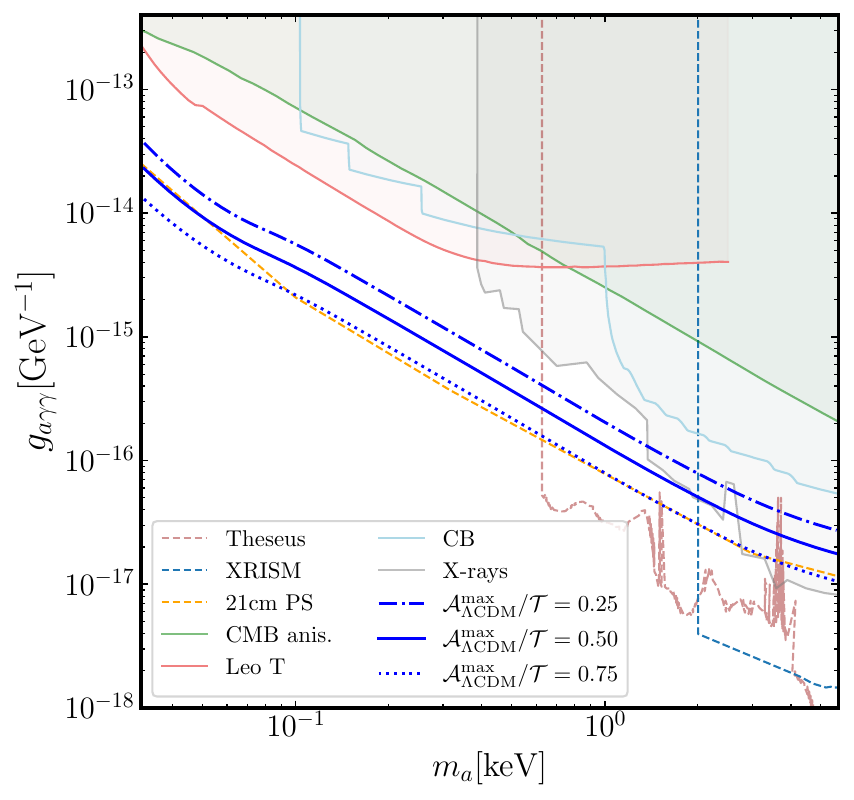}
        \caption{Current (solid lines) and projected (dashed lines) bounds in the $(m_a, g_{a\gamma\gamma})$ plane, assuming that axions constitute the dark matter abundance and remaining agnostic about their production mechanism. The blue lines indicate the projected sensitivity of global 21\,cm detection for three representative benchmark signals.}
        \label{fig:21cmA}
    \end{subfigure}
    \hfill
    \begin{subfigure}[b]{0.45\textwidth}
        \centering
        \includegraphics[width=\textwidth]{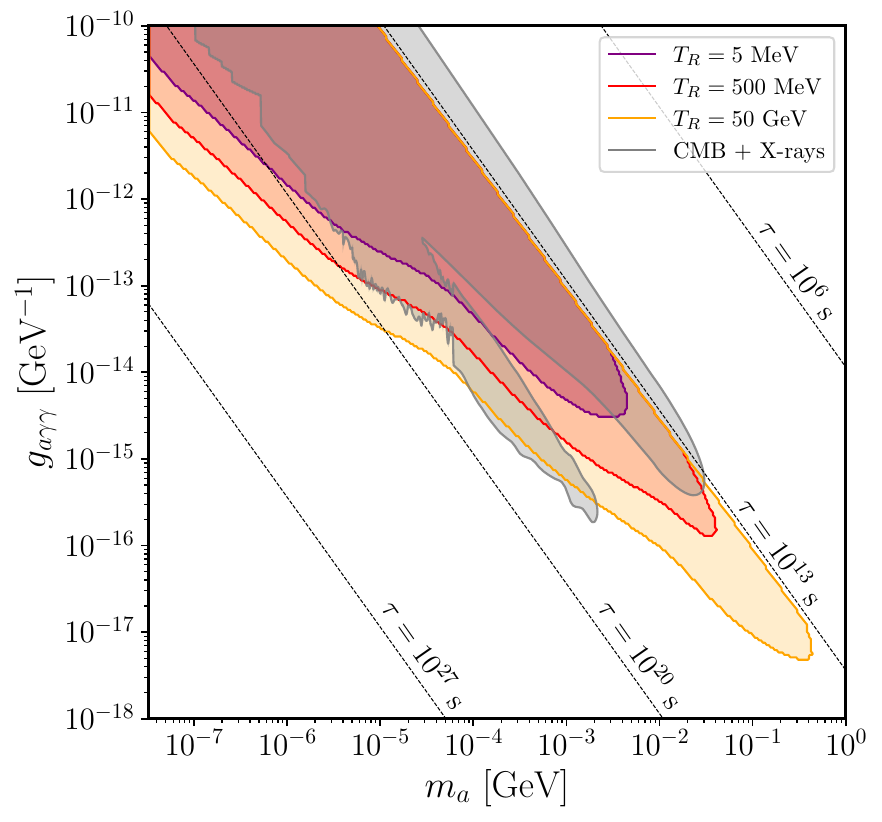}
        \caption{Projected sensitivity to a 21\,cm signal arising from the irreducible axion background for three reheating temperatures. In this case, the axion abundance is evaluated self-consistently. Shaded gray regions are excluded by CMB and X-rays, while solid black lines denote benchmark values of the axion lifetime.}
        \label{fig:21cmB}
    \end{subfigure}
    \caption{Prospects for constraining the parameter space of an axion coupled to photons in the plane spanned by the axion mass $m_a$ and the axion–photon coupling $g_{a\gamma\gamma}$, as defined in Eq.~\eqref{eq:gaggg}. Figures from Ref.~\cite{Cima:2025zmc}.}
    \label{fig:21cm}
\end{figure}

We report here the application of this general model-independent strategy to the case of axions coupled to photons via anomalous dimension 5 interactions. To discuss the results of Ref.~\cite{Cima:2025zmc}, we adopt in this section the same notation as they use and investigate the implications of the dimension-5 axion interaction\footnote{The connection with the Lagrangian in Eq.~\eqref{eq:Laint} is straightforward through the relation $C_\gamma = 2 \pi f_a \, g_{a\gamma\gamma}$. }
\begin{equation}
    \mathcal{L}_{a\gamma\gamma} = \frac{g_{a\gamma\gamma}}{4} \, a F_{\mu\nu} \widetilde{F}^{\mu\nu} \ ,
    \label{eq:gaggg}
\end{equation}
The main findings for this case are reported in Fig.~\ref{fig:21cm}.

We first consider the case in which axions account for the observed dark matter abundance. We do not specify the underlying production mechanism, although we note that freeze-in through the same interaction cannot be responsible for the relic density. Indeed, the requirement of sufficiently small couplings to ensure a long axion lifetime is incompatible with producing the correct abundance via freeze-in. The sensitivity of 21\,cm observations in this scenario is illustrated in Fig.~\ref{fig:21cmA}, where the three blue lines indicate the projected constraints corresponding to a hypothetical detection of the global signal for three benchmark values of the ratio $\mathcal{A}_{\Lambda \rm CDM}^{\rm max}/\mathcal{T}(z)$. In all cases, the signal is assumed to occur at a fixed redshift $z_t = 15$. Also shown in the same figure are current and projected bounds from other probes, allowing us to assess the additional information that a global 21\,cm detection would provide. As can be seen, 21\,cm cosmology has the potential to probe regions of axion parameter space that are currently unexplored. We stress that our results concern the global 21cm signal, while complementary analyses of the 21cm power spectrum, such as Refs.~\cite{Facchinetti:2023slb,Sun:2023acy}, provide additional and independent constraints on the axion parameter space.

The next scenario we consider is non-minimal, in the sense that we allow for the presence of an additional and unspecified cold component that accounts for the observed dark matter abundance. In this setup, the axion represents an additional dark sector state that is produced in the early universe and eventually decays. We evaluate the axion relic abundance self consistently across the relevant parameter space, allowing for freeze-in production via both Primakoff processes and pair annihilations mediated by the axion–photon interaction in Eq.~\eqref{eq:gaggg}. Since this interaction is described by a dimension five operator, the resulting production rate is ultraviolet dominated and therefore sensitive to the reheating temperature $T_R$. To avoid exponentially suppressed Maxwell–Boltzmann rates, we restrict to reheating temperatures satisfying $T_R > m_a$. Moreover, consistency of the effective field theory requires $T_R \lesssim g_{a\gamma\gamma}^{-1}$. We neglect scattering processes involving external $W^\pm$ bosons, which is justified by restricting to reheating temperatures $T_R \leq 50\,\mathrm{GeV}$. The production rate is evaluated both above and below the confinement scale, where perturbative calculations are reliable, and the two regimes are smoothly interpolated. The resulting asymptotic comoving axion number density is given by~\cite{Cima:2025zmc}
\begin{equation}
Y_a^\infty = \left. \frac{n_a(T)}{s(T)}\right|_{T \ll T_R} \simeq
\left( \frac{g_{a\gamma\gamma}}{10^{-15}\,{\rm GeV}} \right)^2
\times
\left\{
\begin{array}{ccl}
1.6 \times 10^{-18} & & T_R = 5\,{\rm MeV} \\
9.0 \times 10^{-17} & & T_R = 500\,{\rm MeV} \\
5.1 \times 10^{-15} & & T_R = 50\,{\rm GeV}
\end{array}
\right. \, .
\end{equation}

The same interaction responsible for their production also mediates the decay $a \rightarrow \gamma\gamma$. These decays inject energy into the intergalactic medium and imprint observable distortions in the 21\,cm signal. The resulting sensitivity for this \emph{irreducible axion} scenario is illustrated in Fig.~\ref{fig:21cmB}. We find that 21\,cm observations can significantly strengthen existing bounds on the irreducible axion abundance for axion masses below a few keV, as well as probe the narrow region of parameter space left open by current X-ray and CMB constraints.

Overall, the results presented in this section demonstrate that axion decays can leave observable imprints in the 21\,cm signal, and that upcoming experiments may access regions of axion parameter space that remain unconstrained by laboratory searches. These findings highlight the potential of 21\,cm cosmology as a novel probe of metastable particles and of the early universe dynamics that govern their production and decay.

\section{Unstable but Harmless: New Axion Portal}
\label{sec:portal}

In this section, we explore the third and final region of the parameter space shown in Fig.~\ref{fig:axion_lifetime_plane}, corresponding to the bottom-right corner where the axion is unstable and decays with a lifetime shorter than one second. In this regime, the axion can neither constitute dark matter nor contribute to dark radiation, but instead acts as a mediator between the SM and a hidden sector. In this portal scenario, the axion provides a dynamical bridge to new degrees of freedom, enabling interactions that would otherwise remain secluded.

Axions as portals have been studied extensively in scenarios where the dark matter candidate is a Dirac or Majorana fermion~\cite{Nomura:2008ru,Gola:2021abm,Bharucha:2022lty,Ghosh:2023tyz,Fitzpatrick:2023xks,Dror:2023fyd,Armando:2023zwz,Allen:2024ndv} or a spin-one vector boson~\cite{Kaneta:2016wvf,Kaneta:2017wfh}. In these cases, the resulting phenomenology depends sensitively on the specific SM fields to which the axion couples, and a wide range of possibilities has been explored in the literature.

Here, we present the recent study of Ref.~\cite{DEramo:2025xef}, which addresses the following question: does the axion portal remain viable when dark matter is a scalar field? This question is subtle and becomes immediately apparent when the problem is examined from an effective field theory perspective. The lowest-order interaction compatible with the axion shift symmetry is a derivative coupling between the axion and the spin-one current of the scalar field,
\begin{equation}
\mathcal{L}_{S \varphi} \supset \mathcal{C}_S \frac{\partial_\mu \varphi}{2 f_\varphi} S^\dagger i \overleftrightarrow{\partial}^\mu S \, ,
\label{eq:ALPortal}
\end{equation}
where $\varphi$ and $S$ denote the axion portal field and the scalar dark matter field, respectively, and the bidirectional derivative is defined as
$S^\dagger \overleftrightarrow{\partial}^\mu S \equiv S^\dagger (\partial^\mu S) - (\partial^\mu S^\dagger) S$.
As is well known from the literature on the QCD axion~\cite{Georgi:1986df}, this interaction can be removed through a field redefinition. For the case of a scalar dark matter field, this transformation takes the explicit form
$S \rightarrow \exp\!\left[i\,\mathcal{C}_S \varphi / (2 f_\varphi)\right] S$.
This observation raises the question of whether the interaction in Eq.~\eqref{eq:ALPortal} leads to any physical consequences at all.\footnote{For QCD axion couplings to the Higgs doublet, the dimension five derivative interaction is redundant and can be removed by a field redefinition. This follows from $SU(2)_L$ invariance, which requires the Higgs doublet to enter the scalar potential only through the gauge invariant combination $H^\dagger H$, which is unaffected by the field redefinition.}

To address this issue, the field redefinition must be applied consistently to the entire Lagrangian, and not only to the operator in Eq.~\eqref{eq:ALPortal}. In particular, it must also act on the scalar potential responsible for stabilizing the dark matter,
$V_S(S) \rightarrow V_S\!\left(\exp\!\left[i\,\mathcal{C}_S \varphi / (2 f_\varphi)\right] S\right)$.
Whether the portal interaction survives therefore depends crucially on the structure of this potential. We are thus led to a particularly interesting conclusion: a deep connection between the symmetry that stabilizes the dark matter field and the resulting phenomenology.

\begin{figure}
    \centering
    \includegraphics[width=\textwidth]{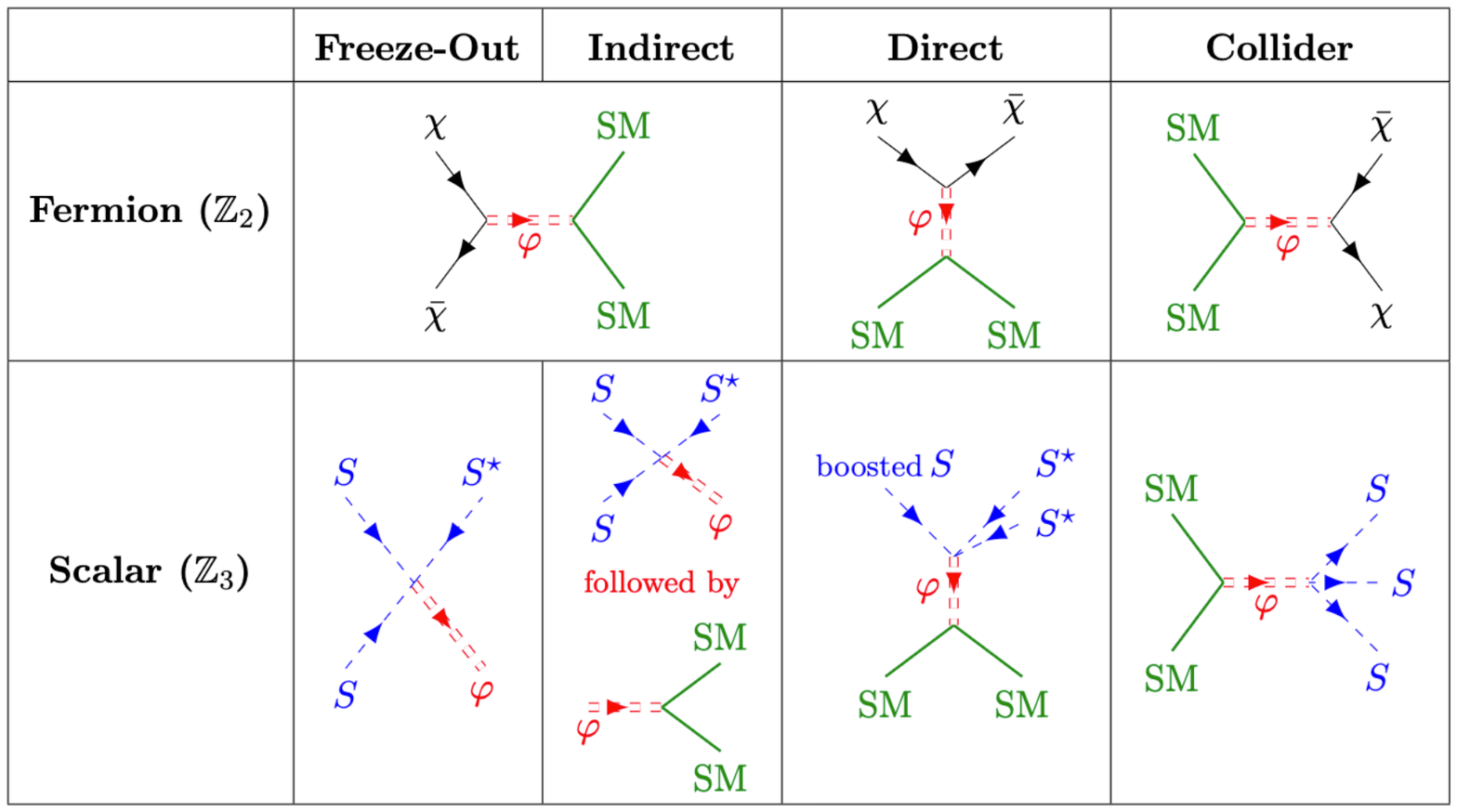}
    \caption{Phenomenology of axion portal scenarios in which the dark matter candidate is either a fermion stabilized by a $\mathbb{Z}_2$ symmetry or a scalar stabilized by a $\mathbb{Z}_3$ symmetry. While scenarios with fermionic dark matter have been extensively explored in the literature, the case of scalar dark matter stabilized by a non-Abelian symmetry exhibits qualitatively new features. Figure from Ref.~\cite{DEramo:2025xef}.}
    \label{fig:ALPortal}
\end{figure}

To explore this, Ref.~\cite{DEramo:2025xef} considered a scenario in which the dark matter field is stabilized by a non-Abelian $\mathbb{Z}_3$ symmetry. In this setup, portal interactions survive due to the presence of cubic terms in the scalar potential, such as $S^3$ and $S^{\dag\,3}$. Such terms would be absent, for example, if the dark matter field were instead stabilized by an Abelian $\mathbb{Z}_2$ symmetry. The resulting phenomenology therefore differs markedly from that of the more familiar fermionic axion portal, as compactly summarized in Fig.~\ref{fig:ALPortal}.

The dark matter abundance is determined by thermal freeze-out through semi-annihilations~\cite{DEramo:2010keq}. Interestingly, when the axion couplings are sufficiently large to ensure thermalization, the relic density becomes independent of the axion coupling to visible-sector particles. Instead, it is controlled solely by the axion–dark matter coupling $\mathcal{C}_S / f_\varphi$ appearing in Eq.~\eqref{eq:ALPortal}, together with the dimensionful coupling $A$ that multiplies the cubic term $S^3$ in the scalar potential. This is the first crucial difference with respect to the more familiar Abelian case with a $\mathbb{Z}_2$ symmetry; it is possible to reproduce the relic density with smaller interactions to SM fields and this helps to be consistent with current axion searches. In other words, the relic density constraint is decoupled from experimental searches today. 

The same semi-annihilation processes can also occur today in astrophysical environments with high dark matter density, allowing one to search for their debris with terrestrial and satellite-based telescopes~\cite{DEramo:2010keq,DEramo:2012fou,Arcadi:2017vis,Queiroz:2019acr}. In this case, the overall indirect detection rate is again insensitive to the axion–SM couplings, since the cosmic rays originate from axions produced on shell through semi-annihilation.\footnote{Annihilations do not contribute to indirect detection spectra since the associated matrix element vanishes once the external dark matter particles are put on-shell.} The produced axions subsequently decay into SM states, and the relative branching ratios determine the resulting indirect detection spectra. In other words, the dark sector dynamics control the overall signal normalization, while the axion–SM couplings shape the spectral features.

Finally, direct detection and collider searches are comparatively less promising avenues for probing this scenario. In the case of direct detection, the reason is straightforward: the amplitude for elastic scattering vanishes once the external dark matter states are taken on-shell. An equivalent and more transparent way to see this is to perform a field redefinition and work in a basis where the axion portal interaction is no longer derivative. In this basis, the dark matter–axion couplings are purely of the form $S^3\varphi$.\footnote{As illustrated in Fig.~\ref{fig:ALPortal}, tree-level direct detection scattering may become possible only if the incoming dark matter particle is sufficiently boosted to overcome the kinematic threshold required to produce two dark matter particles in the final state.}
Similarly, collider searches based on missing energy signatures are less sensitive in this case, since the non-Abelian stabilizing symmetry requires the production of at least three dark matter particles in the final state.

\begin{figure}[t]
    \centering
    \begin{subfigure}[b]{0.45\textwidth}
        \centering
        \includegraphics[width=\textwidth]{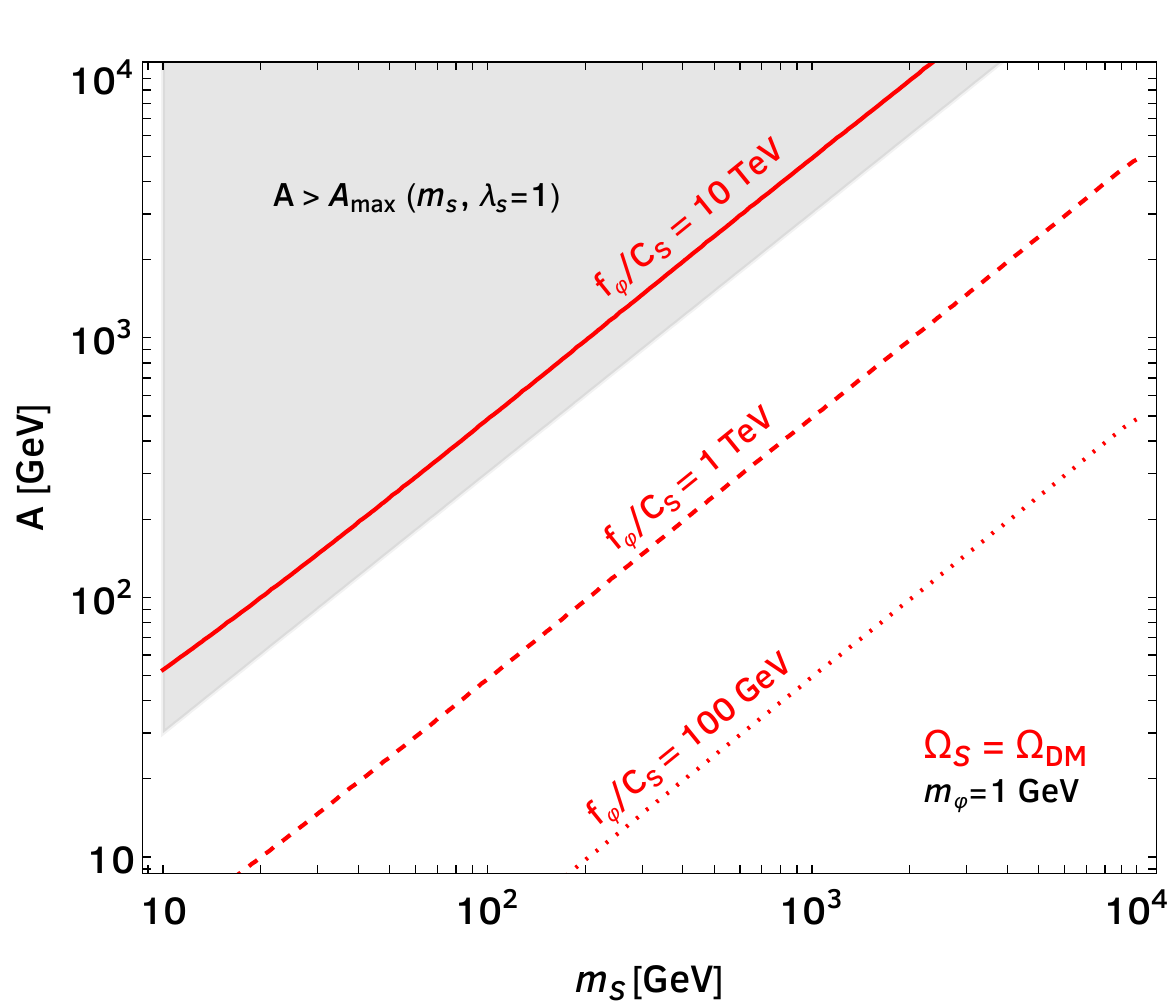}
        \caption{Relic density constraints in slices of parameter space defined in the plane spanned by the dark matter mass $m_S$ and the cubic coupling $A$ in the scalar potential. Different curves correspond to fixed values of the ratio $f_\varphi / \mathcal{C}_S$. The gray shaded region is excluded by theoretical constraints on the scalar potential.}
        \label{fig:ALPortal_DMA}
    \end{subfigure}
    \hfill
    \begin{subfigure}[b]{0.45\textwidth}
        \centering
        \includegraphics[width=\textwidth]{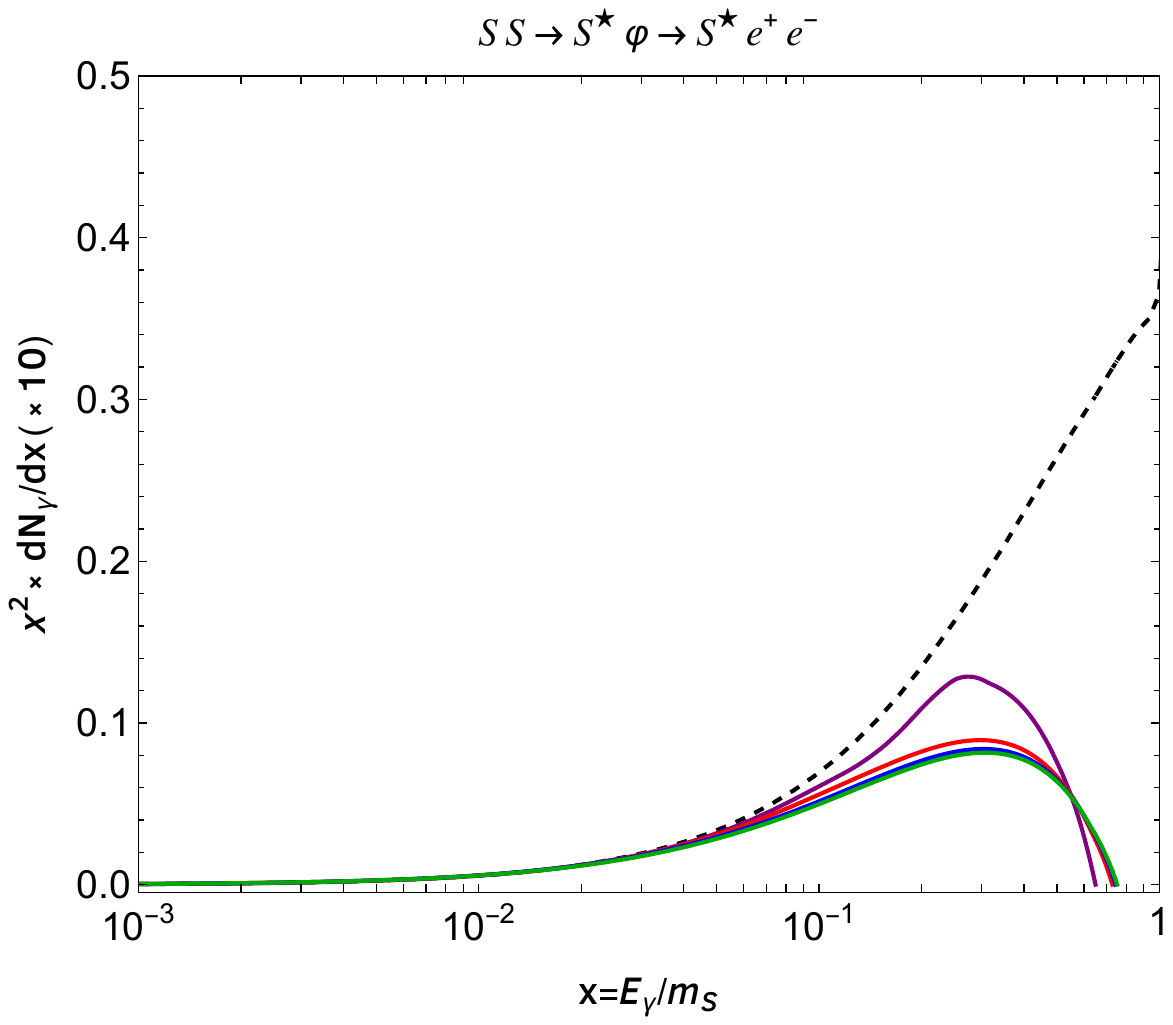}
        \caption{Gamma-ray spectra produced by one-step semi-annihilation processes in which the cascade terminates in an electron–positron pair. The different colored curves correspond to distinct values of $\epsilon_\varphi \equiv m_\varphi / m_S$, while the dashed black lines show the gamma-ray spectra from dark matter annihilation with mass $10\,\mathrm{GeV}$.}
        \label{fig:ALPortal_DMB}
    \end{subfigure}
    \caption{Phenomenology of the axion portal to scalar dark matter when stability is ensured by a non-Abelian $\mathbb{Z}_3$ symmetry. Figures from Ref.~\cite{DEramo:2025xef}.}
    \label{fig:ALPortal_DM}
\end{figure}

The two panels in Fig.~\ref{fig:ALPortal_DM} summarize the dark matter phenomenology. The left panel, Fig.~\ref{fig:ALPortal_DMA}, highlights the regions of parameter space consistent with the observed relic density. As expected, this constraint is independent of the axion--SM couplings, provided the axion attains thermal equilibrium in the early universe. We show representative slices of the $(m_S, A)$ plane, where each curve corresponds to the relic density condition for fixed values of the ratio $f_\varphi / \mathcal{C}_S$. The gray-shaded region is excluded by theoretical constraints on the scalar potential.

The right panel in Fig.~\ref{fig:ALPortal_DMB} illustrates an example of indirect detection observables for a representative benchmark. We consider the case of an axion coupled to electrons and focus on axion masses above the GeV scale, for which the decay $\varphi \rightarrow e^+ e^-$ is kinematically allowed. The relevant process is the semi-annihilation $S S \rightarrow S^\star \varphi$, followed by the subsequent axion decay. This sequence produces gamma rays through final-state radiation, and the resulting injection spectrum can be computed analytically (see App.~A of Ref.~\cite{Mardon:2009rc}). We refer to this mechanism as \emph{one-step semi-annihilation}. Fig.~\ref{fig:ALPortal_DMB} shows the corresponding gamma-ray spectra for different values of the mass ratio $\epsilon_\varphi \equiv m_\varphi / m_S$, illustrating how the spectral shape depends on the axion-to–dark matter mass hierarchy, and compares them with the spectrum expected from standard dark matter annihilation. 

These results highlight distinctive signatures of the axion portal in the scalar dark matter case. While preliminary, they motivate a more systematic and comprehensive analysis aimed at delineating the viable parameter space and identifying the associated observational signatures in greater detail. More broadly, this framework illustrates how axion-like particles, motivated by robust theoretical considerations, can play a role in the dark sector that extends beyond serving as dark matter candidates themselves, acting instead as mediators whose phenomenology is tightly linked to the symmetries stabilizing dark matter. This connection opens new avenues for exploring dark sector dynamics through indirect detection and cosmological observations.

\section{Conclusion}
\label{sec:conclusions}

Let us conclude by stepping back and considering the broader picture that emerges from our analysis. In this work, we have surveyed three broad and complementary regions of axion parameter space, each characterized by a distinct mechanism through which new light degrees of freedom can imprint themselves on cosmological observables. These regimes span scenarios in which axions are effectively stable and contribute to dark radiation or dark matter, cases in which metastable axions decay at cosmologically relevant epochs and leave signatures in the global 21\,cm signal, and finally situations in which short-lived axions act as portals to otherwise hidden sectors. Taken together, these possibilities showcase the remarkable richness of axion phenomenology and the wide variety of cosmological signatures that can arise from axion physics.

While each region relies on its own set of assumptions and exhibits a distinct phenomenology, a clear unifying theme emerges: the early universe constitutes an exceptionally powerful and versatile laboratory for probing axions. The combination of precision cosmological observations, astrophysical measurements, and theoretical consistency requirements enables access to wide and otherwise elusive regions of parameter space, many of which lie far beyond the reach of laboratory-based experiments on Earth.

The results summarized here underscore the extraordinary sensitivity of cosmological observables to light and weakly-coupled particles, and highlight how a remarkable amount of information about axions can be encoded in the thermal and dynamical history of the universe. As observational capabilities continue to advance and theoretical tools become increasingly sophisticated, these cosmological probes are expected to become ever more incisive, enabling progressively stringent tests of axion physics across a wide range of masses and couplings. In this way, the early universe itself emerges as a powerful particle physics laboratory, poised to play a central role in deepening our understanding of physics beyond the SM in the years to come.

\acknowledgments 
We thank the organizers of the Corfu Summer Institute \textit{``Workshop on the Standard Model and Beyond 2025''} for the stimulating atmosphere and for the opportunity to present this work. This research was supported by the Istituto Nazionale di Fisica Nucleare (INFN) through the Theoretical Astroparticle Physics (TAsP) project, and in part by the Italian Ministry of University and Research (MUR) under the Departments of Excellence grant 2023--2027, ``Quantum Frontiers''.

\bibliographystyle{JHEP}
\bibliography{DEramoAxions}

\providecommand{\href}[2]{#2}\begingroup\raggedright\begin{thebibliography}{10}

\bibitem{Peccei:1977hh}
R.D.~Peccei and H.R.~Quinn, \emph{{CP Conservation in the Presence of
  Instantons}}, \href{https://doi.org/10.1103/PhysRevLett.38.1440}{\emph{Phys.
  Rev. Lett.} {\bfseries 38} (1977) 1440}.

\bibitem{Peccei:1977ur}
R.D.~Peccei and H.R.~Quinn, \emph{{Constraints Imposed by CP Conservation in
  the Presence of Instantons}},
  \href{https://doi.org/10.1103/PhysRevD.16.1791}{\emph{Phys. Rev. D}
  {\bfseries 16} (1977) 1791}.

\bibitem{Wilczek:1977pj}
F.~Wilczek, \emph{{Problem of Strong $P$ and $T$ Invariance in the Presence of
  Instantons}}, \href{https://doi.org/10.1103/PhysRevLett.40.279}{\emph{Phys.
  Rev. Lett.} {\bfseries 40} (1978) 279}.

\bibitem{Weinberg:1977ma}
S.~Weinberg, \emph{{A New Light Boson?}},
  \href{https://doi.org/10.1103/PhysRevLett.40.223}{\emph{Phys. Rev. Lett.}
  {\bfseries 40} (1978) 223}.

\bibitem{Georgi:1986df}
H.~Georgi, D.B.~Kaplan and L.~Randall, \emph{{Manifesting the Invisible Axion
  at Low-energies}},
  \href{https://doi.org/10.1016/0370-2693(86)90688-X}{\emph{Phys. Lett. B}
  {\bfseries 169} (1986) 73}.

\bibitem{Langhoff:2022bij}
K.~Langhoff, N.J.~Outmezguine and N.L.~Rodd, \emph{{Irreducible Axion
  Background}},
  \href{https://doi.org/10.1103/PhysRevLett.129.241101}{\emph{Phys. Rev. Lett.}
  {\bfseries 129} (2022) 241101}
  [\href{https://arxiv.org/abs/2209.06216}{{\ttfamily 2209.06216}}].

\bibitem{DEramo:2024lsk}
F.~D'Eramo, A.~Tesi and V.~Vaskonen, \emph{{Irreducible cosmological
  backgrounds of a real scalar with a broken symmetry}},
  \href{https://doi.org/10.1103/PhysRevD.110.095002}{\emph{Phys. Rev. D}
  {\bfseries 110} (2024) 095002}
  [\href{https://arxiv.org/abs/2407.19997}{{\ttfamily 2407.19997}}].

\bibitem{Brust:2013ova}
C.~Brust, D.E.~Kaplan and M.T.~Walters, \emph{{New Light Species and the CMB}},
  \href{https://doi.org/10.1007/JHEP12(2013)058}{\emph{JHEP} {\bfseries 12}
  (2013) 058} [\href{https://arxiv.org/abs/1303.5379}{{\ttfamily 1303.5379}}].

\bibitem{Baumann:2016wac}
D.~Baumann, D.~Green and B.~Wallisch, \emph{{New Target for Cosmic Axion
  Searches}}, \href{https://doi.org/10.1103/PhysRevLett.117.171301}{\emph{Phys.
  Rev. Lett.} {\bfseries 117} (2016) 171301}
  [\href{https://arxiv.org/abs/1604.08614}{{\ttfamily 1604.08614}}].

\bibitem{Ferreira:2018vjj}
R.Z.~Ferreira and A.~Notari, \emph{{Observable Windows for the QCD Axion
  Through the Number of Relativistic Species}},
  \href{https://doi.org/10.1103/PhysRevLett.120.191301}{\emph{Phys. Rev. Lett.}
  {\bfseries 120} (2018) 191301}
  [\href{https://arxiv.org/abs/1801.06090}{{\ttfamily 1801.06090}}].

\bibitem{DEramo:2018vss}
F.~D'Eramo, R.Z.~Ferreira, A.~Notari and J.L.~Bernal, \emph{{Hot Axions and the
  $H_0$ tension}},
  \href{https://doi.org/10.1088/1475-7516/2018/11/014}{\emph{JCAP} {\bfseries
  11} (2018) 014} [\href{https://arxiv.org/abs/1808.07430}{{\ttfamily
  1808.07430}}].

\bibitem{Arias-Aragon:2020qtn}
F.~Arias-Arag{\'o}n, F.~D'eramo, R.Z.~Ferreira, L.~Merlo and A.~Notari,
  \emph{{Cosmic Imprints of XENON1T Axions}},
  \href{https://doi.org/10.1088/1475-7516/2020/11/025}{\emph{JCAP} {\bfseries
  11} (2020) 025} [\href{https://arxiv.org/abs/2007.06579}{{\ttfamily
  2007.06579}}].

\bibitem{Arias-Aragon:2020shv}
F.~Arias-Arag{\'o}n, F.~D'Eramo, R.Z.~Ferreira, L.~Merlo and A.~Notari,
  \emph{{Production of Thermal Axions across the ElectroWeak Phase
  Transition}},
  \href{https://doi.org/10.1088/1475-7516/2021/03/090}{\emph{JCAP} {\bfseries
  03} (2021) 090} [\href{https://arxiv.org/abs/2012.04736}{{\ttfamily
  2012.04736}}].

\bibitem{Green:2021hjh}
D.~Green, Y.~Guo and B.~Wallisch, \emph{{Cosmological implications of
  axion-matter couplings}},
  \href{https://doi.org/10.1088/1475-7516/2022/02/019}{\emph{JCAP} {\bfseries
  02} (2022) 019} [\href{https://arxiv.org/abs/2109.12088}{{\ttfamily
  2109.12088}}].

\bibitem{Yeh:2022heq}
T.-H.~Yeh, J.~Shelton, K.A.~Olive and B.D.~Fields, \emph{{Probing physics
  beyond the standard model: limits from BBN and the CMB independently and
  combined}}, \href{https://doi.org/10.1088/1475-7516/2022/10/046}{\emph{JCAP}
  {\bfseries 10} (2022) 046}
  [\href{https://arxiv.org/abs/2207.13133}{{\ttfamily 2207.13133}}].

\bibitem{Schoneberg:2024ifp}
N.~Sch{\"o}neberg, \emph{{The 2024 BBN baryon abundance update}},
  \href{https://doi.org/10.1088/1475-7516/2024/06/006}{\emph{JCAP} {\bfseries
  06} (2024) 006} [\href{https://arxiv.org/abs/2401.15054}{{\ttfamily
  2401.15054}}].

\bibitem{Planck:2018vyg}
{\scshape Planck} collaboration, \emph{{Planck 2018 results. VI. Cosmological
  parameters}},
  \href{https://doi.org/10.1051/0004-6361/201833910}{\emph{Astron. Astrophys.}
  {\bfseries 641} (2020) A6}
  [\href{https://arxiv.org/abs/1807.06209}{{\ttfamily 1807.06209}}].

\bibitem{ACT:2020gnv}
{\scshape ACT} collaboration, \emph{{The Atacama Cosmology Telescope: DR4 Maps
  and Cosmological Parameters}},
  \href{https://doi.org/10.1088/1475-7516/2020/12/047}{\emph{JCAP} {\bfseries
  12} (2020) 047} [\href{https://arxiv.org/abs/2007.07288}{{\ttfamily
  2007.07288}}].

\bibitem{SimonsObservatory:2018koc}
{\scshape Simons Observatory} collaboration, \emph{{The Simons Observatory:
  Science goals and forecasts}},
  \href{https://doi.org/10.1088/1475-7516/2019/02/056}{\emph{JCAP} {\bfseries
  02} (2019) 056} [\href{https://arxiv.org/abs/1808.07445}{{\ttfamily
  1808.07445}}].

\bibitem{Abazajian:2019eic}
K.~Abazajian et~al., \emph{{CMB-S4 Science Case, Reference Design, and Project
  Plan}},  \href{https://arxiv.org/abs/1907.04473}{{\ttfamily 1907.04473}}.

\bibitem{CMB-S4:2022ght}
{\scshape CMB-S4} collaboration, \emph{{Snowmass 2021 CMB-S4 White Paper}},
  \href{https://arxiv.org/abs/2203.08024}{{\ttfamily 2203.08024}}.

\bibitem{DEramo:2023nzt}
F.~D'Eramo, F.~Hajkarim and A.~Lenoci, \emph{{Dark radiation from the
  primordial thermal bath in momentum space}},
  \href{https://doi.org/10.1088/1475-7516/2024/03/009}{\emph{JCAP} {\bfseries
  03} (2024) 009} [\href{https://arxiv.org/abs/2311.04974}{{\ttfamily
  2311.04974}}].

\bibitem{DEramo:2024jhn}
F.~D'Eramo and A.~Lenoci, \emph{{Back to the phase space: Thermal axion dark
  radiation via couplings to standard model fermions}},
  \href{https://doi.org/10.1103/PhysRevD.110.116028}{\emph{Phys. Rev. D}
  {\bfseries 110} (2024) 116028}
  [\href{https://arxiv.org/abs/2410.21253}{{\ttfamily 2410.21253}}].

\bibitem{DEramo:2021usm}
F.~D'Eramo and S.~Yun, \emph{{Flavor violating axions in the early Universe}},
  \href{https://doi.org/10.1103/PhysRevD.105.075002}{\emph{Phys. Rev. D}
  {\bfseries 105} (2022) 075002}
  [\href{https://arxiv.org/abs/2111.12108}{{\ttfamily 2111.12108}}].

\bibitem{Badziak:2024qjg}
M.~Badziak and M.~Laletin, \emph{{Precise predictions for the QCD axion
  contribution to dark radiation with full phase-space evolution}},
  \href{https://doi.org/10.1007/JHEP02(2025)108}{\emph{JHEP} {\bfseries 02}
  (2025) 108} [\href{https://arxiv.org/abs/2410.18186}{{\ttfamily
  2410.18186}}].

\bibitem{Graf:2010tv}
P.~Graf and F.D.~Steffen, \emph{{Thermal axion production in the primordial
  quark-gluon plasma}},
  \href{https://doi.org/10.1103/PhysRevD.83.075011}{\emph{Phys. Rev. D}
  {\bfseries 83} (2011) 075011}
  [\href{https://arxiv.org/abs/1008.4528}{{\ttfamily 1008.4528}}].

\bibitem{Salvio:2013iaa}
A.~Salvio, A.~Strumia and W.~Xue, \emph{{Thermal axion production}},
  \href{https://doi.org/10.1088/1475-7516/2014/01/011}{\emph{JCAP} {\bfseries
  01} (2014) 011} [\href{https://arxiv.org/abs/1310.6982}{{\ttfamily
  1310.6982}}].

\bibitem{DEramo:2021psx}
F.~D'Eramo, F.~Hajkarim and S.~Yun, \emph{{Thermal Axion Production at Low
  Temperatures: A Smooth Treatment of the QCD Phase Transition}},
  \href{https://doi.org/10.1103/PhysRevLett.128.152001}{\emph{Phys. Rev. Lett.}
  {\bfseries 128} (2022) 152001}
  [\href{https://arxiv.org/abs/2108.04259}{{\ttfamily 2108.04259}}].

\bibitem{DEramo:2021lgb}
F.~D'Eramo, F.~Hajkarim and S.~Yun, \emph{{Thermal QCD Axions across
  Thresholds}}, \href{https://doi.org/10.1007/JHEP10(2021)224}{\emph{JHEP}
  {\bfseries 10} (2021) 224}
  [\href{https://arxiv.org/abs/2108.05371}{{\ttfamily 2108.05371}}].

\bibitem{Bouzoud:2024bom}
K.~Bouzoud and J.~Ghiglieri, \emph{{Thermal axion production at hard and soft
  momenta}}, \href{https://doi.org/10.1007/JHEP01(2025)163}{\emph{JHEP}
  {\bfseries 01} (2025) 163}
  [\href{https://arxiv.org/abs/2404.06113}{{\ttfamily 2404.06113}}].

\bibitem{Ferreira:2020bpb}
R.Z.~Ferreira, A.~Notari and F.~Rompineve,
  \emph{{Dine-Fischler-Srednicki-Zhitnitsky axion in the CMB}},
  \href{https://doi.org/10.1103/PhysRevD.103.063524}{\emph{Phys. Rev. D}
  {\bfseries 103} (2021) 063524}
  [\href{https://arxiv.org/abs/2012.06566}{{\ttfamily 2012.06566}}].

\bibitem{Notari:2022ffe}
A.~Notari, F.~Rompineve and G.~Villadoro, \emph{{Improved Hot Dark Matter Bound
  on the QCD Axion}},
  \href{https://doi.org/10.1103/PhysRevLett.131.011004}{\emph{Phys. Rev. Lett.}
  {\bfseries 131} (2023) 011004}
  [\href{https://arxiv.org/abs/2211.03799}{{\ttfamily 2211.03799}}].

\bibitem{Bianchini:2023ubu}
F.~Bianchini, G.G.~di~Cortona and M.~Valli, \emph{{QCD axion: Some like it
  hot}}, \href{https://doi.org/10.1103/PhysRevD.110.123527}{\emph{Phys. Rev. D}
  {\bfseries 110} (2024) 123527}
  [\href{https://arxiv.org/abs/2310.08169}{{\ttfamily 2310.08169}}].

\bibitem{Caloni:2022uya}
L.~Caloni, M.~Gerbino, M.~Lattanzi and L.~Visinelli, \emph{{Novel cosmological
  bounds on thermally-produced axion-like particles}},
  \href{https://doi.org/10.1088/1475-7516/2022/09/021}{\emph{JCAP} {\bfseries
  09} (2022) 021} [\href{https://arxiv.org/abs/2205.01637}{{\ttfamily
  2205.01637}}].

\bibitem{Badziak:2025mkt}
M.~Badziak, A.~Gomu{\l}ka, M.~Laletin and K.~Szafra{\'n}ski, \emph{{Improved
  cosmological constraints on axion-lepton interactions}},
  \href{https://arxiv.org/abs/2511.14864}{{\ttfamily 2511.14864}}.

\bibitem{DEramo:2022nvb}
F.~D'Eramo, E.~Di~Valentino, W.~Giar{\`e}, F.~Hajkarim, A.~Melchiorri, O.~Mena
  et~al., \emph{{Cosmological bound on the QCD axion mass, redux}},
  \href{https://doi.org/10.1088/1475-7516/2022/09/022}{\emph{JCAP} {\bfseries
  09} (2022) 022} [\href{https://arxiv.org/abs/2205.07849}{{\ttfamily
  2205.07849}}].

\bibitem{Cowsik:1972gh}
R.~Cowsik and J.~McClelland, \emph{{An Upper Limit on the Neutrino Rest Mass}},
  \href{https://doi.org/10.1103/PhysRevLett.29.669}{\emph{Phys. Rev. Lett.}
  {\bfseries 29} (1972) 669}.

\bibitem{Hall:2009bx}
L.J.~Hall, K.~Jedamzik, J.~March-Russell and S.M.~West, \emph{{Freeze-In
  Production of FIMP Dark Matter}},
  \href{https://doi.org/10.1007/JHEP03(2010)080}{\emph{JHEP} {\bfseries 03}
  (2010) 080} [\href{https://arxiv.org/abs/0911.1120}{{\ttfamily 0911.1120}}].

\bibitem{Co:2015pka}
R.T.~Co, F.~D'Eramo, L.J.~Hall and D.~Pappadopulo, \emph{{Freeze-In Dark Matter
  with Displaced Signatures at Colliders}},
  \href{https://doi.org/10.1088/1475-7516/2015/12/024}{\emph{JCAP} {\bfseries
  12} (2015) 024} [\href{https://arxiv.org/abs/1506.07532}{{\ttfamily
  1506.07532}}].

\bibitem{Evans:2016zau}
J.A.~Evans and J.~Shelton, \emph{{Long-Lived Staus and Displaced Leptons at the
  LHC}}, \href{https://doi.org/10.1007/JHEP04(2016)056}{\emph{JHEP} {\bfseries
  04} (2016) 056} [\href{https://arxiv.org/abs/1601.01326}{{\ttfamily
  1601.01326}}].

\bibitem{DEramo:2017ecx}
F.~D'Eramo, N.~Fernandez and S.~Profumo, \emph{{Dark Matter Freeze-in
  Production in Fast-Expanding Universes}},
  \href{https://doi.org/10.1088/1475-7516/2018/02/046}{\emph{JCAP} {\bfseries
  02} (2018) 046} [\href{https://arxiv.org/abs/1712.07453}{{\ttfamily
  1712.07453}}].

\bibitem{Calibbi:2018fqf}
L.~Calibbi, L.~Lopez-Honorez, S.~Lowette and A.~Mariotti,
  \emph{{Singlet-Doublet Dark Matter Freeze-in: LHC displaced signatures versus
  cosmology}}, \href{https://doi.org/10.1007/JHEP09(2018)037}{\emph{JHEP}
  {\bfseries 09} (2018) 037}
  [\href{https://arxiv.org/abs/1805.04423}{{\ttfamily 1805.04423}}].

\bibitem{Belanger:2018sti}
G.~B{\'e}langer et~al., \emph{{LHC-friendly minimal freeze-in models}},
  \href{https://doi.org/10.1007/JHEP02(2019)186}{\emph{JHEP} {\bfseries 02}
  (2019) 186} [\href{https://arxiv.org/abs/1811.05478}{{\ttfamily
  1811.05478}}].

\bibitem{Bae:2020dwf}
K.J.~Bae, M.~Park and M.~Zhang, \emph{{Demystifying freeze-in dark matter at
  the LHC}}, \href{https://doi.org/10.1103/PhysRevD.101.115036}{\emph{Phys.
  Rev. D} {\bfseries 101} (2020) 115036}
  [\href{https://arxiv.org/abs/2001.02142}{{\ttfamily 2001.02142}}].

\bibitem{Calibbi:2021fld}
L.~Calibbi, F.~D'Eramo, S.~Junius, L.~Lopez-Honorez and A.~Mariotti,
  \emph{{Displaced new physics at colliders and the early universe before its
  first second}}, \href{https://doi.org/10.1007/JHEP05(2021)234}{\emph{JHEP}
  {\bfseries 05} (2021) 234}
  [\href{https://arxiv.org/abs/2102.06221}{{\ttfamily 2102.06221}}].

\bibitem{McDonald:2015ljz}
J.~McDonald, \emph{{Warm Dark Matter via Ultra-Violet Freeze-In: Reheating
  Temperature and Non-Thermal Distribution for Fermionic Higgs Portal Dark
  Matter}}, \href{https://doi.org/10.1088/1475-7516/2016/08/035}{\emph{JCAP}
  {\bfseries 08} (2016) 035}
  [\href{https://arxiv.org/abs/1512.06422}{{\ttfamily 1512.06422}}].

\bibitem{Roland:2016gli}
S.B.~Roland and B.~Shakya, \emph{{Cosmological Imprints of Frozen-In Light
  Sterile Neutrinos}},
  \href{https://doi.org/10.1088/1475-7516/2017/05/027}{\emph{JCAP} {\bfseries
  05} (2017) 027} [\href{https://arxiv.org/abs/1609.06739}{{\ttfamily
  1609.06739}}].

\bibitem{Heeck:2017xbu}
J.~Heeck and D.~Teresi, \emph{{Cold keV dark matter from decays and
  scatterings}}, \href{https://doi.org/10.1103/PhysRevD.96.035018}{\emph{Phys.
  Rev. D} {\bfseries 96} (2017) 035018}
  [\href{https://arxiv.org/abs/1706.09909}{{\ttfamily 1706.09909}}].

\bibitem{Bae:2017dpt}
K.J.~Bae, A.~Kamada, S.P.~Liew and K.~Yanagi, \emph{{Light axinos from
  freeze-in: production processes, phase space distributions, and Ly-$\alpha$
  forest constraints}},
  \href{https://doi.org/10.1088/1475-7516/2018/01/054}{\emph{JCAP} {\bfseries
  01} (2018) 054} [\href{https://arxiv.org/abs/1707.06418}{{\ttfamily
  1707.06418}}].

\bibitem{Kamada:2019kpe}
A.~Kamada and K.~Yanagi, \emph{{Constraining FIMP from the structure formation
  of the Universe: analytic mapping from $m_{WDM}$}},
  \href{https://doi.org/10.1088/1475-7516/2019/11/029}{\emph{JCAP} {\bfseries
  11} (2019) 029} [\href{https://arxiv.org/abs/1907.04558}{{\ttfamily
  1907.04558}}].

\bibitem{DEramo:2020gpr}
F.~D'Eramo and A.~Lenoci, \emph{{Lower mass bounds on FIMP dark matter produced
  via freeze-in}},
  \href{https://doi.org/10.1088/1475-7516/2021/10/045}{\emph{JCAP} {\bfseries
  10} (2021) 045} [\href{https://arxiv.org/abs/2012.01446}{{\ttfamily
  2012.01446}}].

\bibitem{DEramo:2025jsb}
F.~D'Eramo, A.~Lenoci and A.~Dekker, \emph{{Dark matter freeze-in and
  small-scale observables: Novel mass bounds and viable particle candidates}},
  \href{https://doi.org/10.1103/j62q-cvkr}{\emph{Phys. Rev. D} {\bfseries 112}
  (2025) 116008} [\href{https://arxiv.org/abs/2506.13864}{{\ttfamily
  2506.13864}}].

\bibitem{Baumholzer:2020hvx}
S.~Baumholzer, V.~Brdar and E.~Morgante, \emph{{Structure Formation Limits on
  Axion-Like Dark Matter}},
  \href{https://doi.org/10.1088/1475-7516/2021/05/004}{\emph{JCAP} {\bfseries
  05} (2021) 004} [\href{https://arxiv.org/abs/2012.09181}{{\ttfamily
  2012.09181}}].

\bibitem{Becker:2025yvb}
M.~Becker, J.~Harz, E.~Morgante, C.~Puchades-Ib{\'a}{\~n}ez and P.~Schwaller,
  \emph{{ALP production from abelian gauge bosons: beyond hard thermal loops}},
  \href{https://doi.org/10.1007/JHEP06(2025)160}{\emph{JHEP} {\bfseries 06}
  (2025) 160} [\href{https://arxiv.org/abs/2502.01729}{{\ttfamily
  2502.01729}}].

\bibitem{Cima:2025zmc}
F.~Cima and F.~D'Eramo, \emph{{Probing non-minimal dark sectors via the 21 cm
  line at cosmic dawn}},
  \href{https://doi.org/10.1088/1475-7516/2026/02/020}{\emph{JCAP} {\bfseries
  02} (2026) 020} [\href{https://arxiv.org/abs/2507.10664}{{\ttfamily
  2507.10664}}].

\bibitem{Facchinetti:2023slb}
G.~Facchinetti, L.~Lopez-Honorez, Y.~Qin and A.~Mesinger, \emph{{21cm signal
  sensitivity to dark matter decay}},
  \href{https://doi.org/10.1088/1475-7516/2024/01/005}{\emph{JCAP} {\bfseries
  01} (2024) 005} [\href{https://arxiv.org/abs/2308.16656}{{\ttfamily
  2308.16656}}].

\bibitem{Sun:2023acy}
Y.~Sun, J.W.~Foster, H.~Liu, J.B.~Mu{\~n}oz and T.R.~Slatyer,
  \emph{{Inhomogeneous energy injection in the 21-cm power spectrum:
  Sensitivity to dark matter decay}},
  \href{https://doi.org/10.1103/PhysRevD.111.043015}{\emph{Phys. Rev. D}
  {\bfseries 111} (2025) 043015}
  [\href{https://arxiv.org/abs/2312.11608}{{\ttfamily 2312.11608}}].

\bibitem{Nomura:2008ru}
Y.~Nomura and J.~Thaler, \emph{{Dark Matter through the Axion Portal}},
  \href{https://doi.org/10.1103/PhysRevD.79.075008}{\emph{Phys. Rev. D}
  {\bfseries 79} (2009) 075008}
  [\href{https://arxiv.org/abs/0810.5397}{{\ttfamily 0810.5397}}].

\bibitem{Gola:2021abm}
S.~Gola, S.~Mandal and N.~Sinha, \emph{{ALP-portal majorana dark matter}},
  \href{https://doi.org/10.1142/S0217751X22501317}{\emph{Int. J. Mod. Phys. A}
  {\bfseries 37} (2022) 2250131}
  [\href{https://arxiv.org/abs/2106.00547}{{\ttfamily 2106.00547}}].

\bibitem{Bharucha:2022lty}
A.~Bharucha, F.~Br{\"u}mmer, N.~Desai and S.~Mutzel, \emph{{Axion-like
  particles as mediators for dark matter: beyond freeze-out}},
  \href{https://doi.org/10.1007/JHEP02(2023)141}{\emph{JHEP} {\bfseries 02}
  (2023) 141} [\href{https://arxiv.org/abs/2209.03932}{{\ttfamily
  2209.03932}}].

\bibitem{Ghosh:2023tyz}
D.K.~Ghosh, A.~Ghoshal and S.~Jeesun, \emph{{Axion-like particle (ALP) portal
  freeze-in dark matter confronting ALP search experiments}},
  \href{https://doi.org/10.1007/JHEP01(2024)026}{\emph{JHEP} {\bfseries 01}
  (2024) 026} [\href{https://arxiv.org/abs/2305.09188}{{\ttfamily
  2305.09188}}].

\bibitem{Fitzpatrick:2023xks}
P.J.~Fitzpatrick, Y.~Hochberg, E.~Kuflik, R.~Ovadia and Y.~Soreq, \emph{{Dark
  matter through the axion-gluon portal}},
  \href{https://doi.org/10.1103/PhysRevD.108.075003}{\emph{Phys. Rev. D}
  {\bfseries 108} (2023) 075003}
  [\href{https://arxiv.org/abs/2306.03128}{{\ttfamily 2306.03128}}].

\bibitem{Dror:2023fyd}
J.A.~Dror, S.~Gori and P.~Munbodh, \emph{{QCD axion-mediated dark matter}},
  \href{https://doi.org/10.1007/JHEP09(2023)128}{\emph{JHEP} {\bfseries 09}
  (2023) 128} [\href{https://arxiv.org/abs/2306.03145}{{\ttfamily
  2306.03145}}].

\bibitem{Armando:2023zwz}
G.~Armando, P.~Panci, J.~Weiss and R.~Ziegler, \emph{{Leptonic ALP portal to
  the dark sector}},
  \href{https://doi.org/10.1103/PhysRevD.109.055029}{\emph{Phys. Rev. D}
  {\bfseries 109} (2024) 055029}
  [\href{https://arxiv.org/abs/2310.05827}{{\ttfamily 2310.05827}}].

\bibitem{Allen:2024ndv}
S.~Allen, A.~Blackburn, O.~Cardenas, Z.~Messenger, N.H.~Nguyen and B.~Shuve,
  \emph{{Electroweak axion portal to dark matter}},
  \href{https://doi.org/10.1103/PhysRevD.110.095010}{\emph{Phys. Rev. D}
  {\bfseries 110} (2024) 095010}
  [\href{https://arxiv.org/abs/2405.02403}{{\ttfamily 2405.02403}}].

\bibitem{Kaneta:2016wvf}
K.~Kaneta, H.-S.~Lee and S.~Yun, \emph{{Portal Connecting Dark Photons and
  Axions}}, \href{https://doi.org/10.1103/PhysRevLett.118.101802}{\emph{Phys.
  Rev. Lett.} {\bfseries 118} (2017) 101802}
  [\href{https://arxiv.org/abs/1611.01466}{{\ttfamily 1611.01466}}].

\bibitem{Kaneta:2017wfh}
K.~Kaneta, H.-S.~Lee and S.~Yun, \emph{{Dark photon relic dark matter
  production through the dark axion portal}},
  \href{https://doi.org/10.1103/PhysRevD.95.115032}{\emph{Phys. Rev. D}
  {\bfseries 95} (2017) 115032}
  [\href{https://arxiv.org/abs/1704.07542}{{\ttfamily 1704.07542}}].

\bibitem{DEramo:2025xef}
F.~D'Eramo and T.~Sassi, \emph{{Axion portal to scalar Dark Matter: unveiling
  stabilizing symmetry footprints}},
  \href{https://doi.org/10.1007/JHEP07(2025)031}{\emph{JHEP} {\bfseries 07}
  (2025) 031} [\href{https://arxiv.org/abs/2502.19491}{{\ttfamily
  2502.19491}}].

\bibitem{DEramo:2010keq}
F.~D'Eramo and J.~Thaler, \emph{{Semi-annihilation of Dark Matter}},
  \href{https://doi.org/10.1007/JHEP06(2010)109}{\emph{JHEP} {\bfseries 06}
  (2010) 109} [\href{https://arxiv.org/abs/1003.5912}{{\ttfamily 1003.5912}}].

\bibitem{DEramo:2012fou}
F.~D'Eramo, M.~McCullough and J.~Thaler, \emph{{Multiple Gamma Lines from
  Semi-Annihilation}},
  \href{https://doi.org/10.1088/1475-7516/2013/04/030}{\emph{JCAP} {\bfseries
  04} (2013) 030} [\href{https://arxiv.org/abs/1210.7817}{{\ttfamily
  1210.7817}}].

\bibitem{Arcadi:2017vis}
G.~Arcadi, F.S.~Queiroz and C.~Siqueira, \emph{{The Semi-Hooperon: Gamma-ray
  and anti-proton excesses in the Galactic Center}},
  \href{https://doi.org/10.1016/j.physletb.2017.10.065}{\emph{Phys. Lett. B}
  {\bfseries 775} (2017) 196}
  [\href{https://arxiv.org/abs/1706.02336}{{\ttfamily 1706.02336}}].

\bibitem{Queiroz:2019acr}
F.S.~Queiroz and C.~Siqueira, \emph{{Search for Semi-Annihilating Dark Matter
  with Fermi-LAT, H.E.S.S., Planck, and the Cherenkov Telescope Array}},
  \href{https://doi.org/10.1088/1475-7516/2019/04/048}{\emph{JCAP} {\bfseries
  04} (2019) 048} [\href{https://arxiv.org/abs/1901.10494}{{\ttfamily
  1901.10494}}].

\bibitem{Mardon:2009rc}
J.~Mardon, Y.~Nomura, D.~Stolarski and J.~Thaler, \emph{{Dark Matter Signals
  from Cascade Annihilations}},
  \href{https://doi.org/10.1088/1475-7516/2009/05/016}{\emph{JCAP} {\bfseries
  05} (2009) 016} [\href{https://arxiv.org/abs/0901.2926}{{\ttfamily
  0901.2926}}].

\end{thebibliography}\endgroup

\end{document}